\documentclass[conference]{IEEEtran}
\usepackage{flushend}
\usepackage{graphicx}
\usepackage{url}
\usepackage{tabularx}
\usepackage{color}
\usepackage{comment}
\usepackage{cite}

\begin{document}

\author{\IEEEauthorblockN{Joerg Evermann}
\IEEEauthorblockA{Faculty of Business Administration \\
Memorial University of Newfoundland \\
St. John's, NL, Canada \\
Email: jevermann@mun.ca}}

\title{Adapting Workflow Management Systems to BFT Blockchains -- The YAWL Example}

\maketitle

\begin{abstract}
Blockchain technology provides an auditable and tamper-proof distributed storage infrastructure for information records. This can be leveraged to support distributed workflow management. Compared to proof-of-work consensus, popularized by Bitcoin and Ethereum, blockchains based on BFT (byzantine fault tolerance) ordering consensus trade scalability for immediacy and finality of consensus. This makes them easier to use as distribution infrastructure, as applications need not be adapted to deal with eventual consistency and delayed consensus of proof-of-work blockchains. Hence, applications such as workflow engines can be easily ported to such a blockchain infrastructure to take advantage of their decentralized integrity assurance and information distribution model. In this paper we describe how the YAWL workflow engine can be used on a BFT based blockchain infrastructure to enable collaborative workflows across different organizations.  
\end{abstract}


\section{Introduction}

Inter-organizational business processes may include participants in adversarial relationships that have to jointly execute business processes. Trust in the state of a process instance and in the correct execution of activities by other stakeholders may be lacking. Blockchain technology can provide a trusted, distributed workflow execution infrastructure for such situations.

A blockchain cryptographically signs a series of blocks, containing transactions, so that it is difficult or impossible to alter earlier blocks in the chain without this being detectable. In a distributed blockchain, actors independently order and validate transactions, add them to the blockchain, and replicate the chain across nodes. Actors must reach agreement regarding the order and validity of transactions and blocks. In workflow execution, it is important that actors agree on the ''state of work'' as this determines the set of next valid activities in the process. Hence, it is natural to use blockchain transactions to describe the state of work. 

Blockchain technology admits different system designs, and workflow management systems (WfMS) can be implemented in different ways on different types of blockchains. In this paper we present a novel architecture for blockchain-based workflow management. Our contributions are twofold:

First, in contrast to prior work, which has focused on proof-of-work blockchains, we show that a blockchain based on a Byzantine Fault Tolerance (BFT) ordering consensus protocol can be used as workflow execution infrastructure.

Second, in contrast to earlier work, we port an existing, full-featured workflow engine to a blockchain-based workflow management system (WfMS) without smart contracts. The particular workflow system we focus on here is the YAWL system \cite{ter2009modern}, chosen because of its open source implementation and its complete coverage of workflow patterns.

Our approach is independent of any particular workflow language and workflow language semantics as we do not implement a BPMN or workflow net model processor but instead focus on events in work item lifecycles. This language independence allows an easy extension to other workflow engines, as long they support similar work item lifecycles. 

The remainder of the paper is structured as follows. Section~\ref{sec:related} reviews related work on blockchain-based WfMS. We then describe the principles of distributed blockchains with a focus on BFT-based consensus (Sec.~\ref{sec:blockchains}). Section~\ref{sec:architecture} describes the architecture of our system. Section~\ref{sec:prototype} presents our adaptation of the YAWL system\footnote{Source code is available from \url{https://joerg.evermann.ca/software.html}}. The final Sec.~\ref{sec:discussion} discusses corectness guarantees of our architecture, its limitations, and future work.

\section{Related Work}
\label{sec:related}

Recent literature presents a number of workflow execution prototypes implementations, all of them using smart contracts. A smart contract is software code that is stored on the blockchain and execuated as part of blockchain transaction processing. In this way, a smart contract ensures code integrity and consensus on execution results. Many blockchains provide a language and virtual machine for smart contract execution, such as the Solidity language and associated virtual machine originally developed for the Ethereum blockchain.

In a project driven by a financial institution, a domain-specific workflow implementation using Ethereum and smart contracts supports the digital document flow in the import/export trading domain \cite{fridgen2017implementation,fridgen2018cross}. The project demonstrates lowered execution cost, and claims increased transparency and trust among trading partners. Another domain-specific blockchain-based workflow project in the real-estate domain also uses Ethereum and smart contracts \cite{hukkinen2017distributed}. The authors claim the lack of a central agency will make it difficult for regulators to enforce obligations and responsibilities of trading partners.

A domain-agnostic blockchain-based workflow execution system \cite{weber2016using,weber2016untrusted} uses Ethereum smart contracts as choreography monitors, where they monitors execution status and validity of workflow messages, or as active mediators, where they ''drive'' the process by sending and receiving messages\footnote{An Ethereum transaction is a message that is signed by an externally-owned account using its public key. Smart contracts can call other contracts by sending messages but as they do not have a public/private key pair for signing messages, those messages are not considered transactions.} according to a process model. BPMN models are translated into smart contracts. Nodes monitor the blockchain for messages from the smart contract and create transactions for the smart contract. The cost for executing smart contracts and the execution latency are recognized as limitations. A comparison between the public Ethereum blockchain and the Amazon Simple Workflow Service shows blockchain-based execution costs to be two orders of magnitude higher \cite{rimba2017comparing,rimba2018quantifying}. Hence, optimizing the space and computational requirements for smart contracts is important \cite{garcia2017optimized}. BPMN models are first translated to Petri Nets \cite{dijkman2008semantics}, to which existing minimizing algorithms are applied. The minimized Petri nets are then compiled into smart contracts, achieving up to 25\% reduction in execution costs \cite{weber2016using,weber2016untrusted}, while also improving throughput time. Building on lessons learned from \cite{weber2016using,weber2016untrusted}, Caterpillar is an open-source blockchain-based workflow management system \cite{lopez2017caterpillar}. Developed using the Node.js JavaScript runtime it uses the Solidity compiler solc and the Ethereum client geth to provide a distributed execution environment for BPMN-based process models. The Caterpillar system has been extended to directly interpret BPMN models, i.e. it provides a workflow engine as a set of smart contracts. Lorikeet \cite{DBLP:journals/insk/CiccioCDGLLMPTW19} is similar to the original Caterpillar system, also based on BPMN models that are translated to smart contracts.

While most implementations use a flow-based workflow specification, declarative workflows can also be deployed on a blockchain infrastructure \cite{madsen2018collaboration}. This approach is also implemented on Ethereum using Solidity smart contracts.

In summary, while existing work varies in terms of features and capabilities, \emph{all existing blockchain-based workflow execution systems are based on proof-of-work blockchains} and \emph{all use smart contracts}. More specifically, \emph{all are based on the Ethereum blockchain} and its smart contract virtual machine. Moreover, \emph{all (re-)implement a workflow engine}, for example for BPMN, in smart contracts. In contrast, our work is based neither on proof-of-work chains, nor on smart contracts, nor do we need to implement or re-implement a workflow engine.

\section{Blockchains}
\label{sec:blockchains}

A blockchain records transactions in consecutive blocks. A transaction can be any kind of content. Integrity is maintained by hashing the content of each block, which also contains the hash of the previous block. Hence, altering a block requires changing all following blocks. In a typical distributed blockchain, blocks are replicated across nodes. New transactions may originate on any node and must be recorded in new blocks. The key challenge is to achieve consensus on the validity and order of transactions and blocks, despite nodes that are characterized by ''byzantine faults'': they may not respond correctly, may respond unpredictably, or may become altogether unresponsive.

\subsection{Proof-of-Work Consensus}

Bitcoin popularized proof-of-work for consensus finding and securing the blockchain. New transactions are distributed to all nodes, independently validated, and added to a node's transaction pool. A node can independently propose new blocks from transactions in its pool, based on its latest block and its hash, and distribute the new blocks to other nodes. Depending on network speeds and topology, nodes may have different sets of blocks and transactions, and hence may propose different blocks, leading to \emph{side branches}. Node considers the longest branch as their current main branch and propose new blocks based on it. Transactions in side branches are not considered valid. When a side branch becomes longer than the current main branch, the chain undergoes a \emph{reorganization}: What was the side branch is validated and becomes the main branch. What was the main branch is considered invalid and becomes a side branch. Transactions no longer in the main branch are added back to the transaction pool to be included in future blocks. Hence, different nodes may consider different blocks and transactions as valid. As proposed blocks are distributed across the network, nodes will eventually reach a consensus regarding valid blocks and transactions, and their order in the main branch of the chain.

To limit the rate of new block proposals and to secure the blockchain against attacks, block proposers must solve a hard problem (''proof-of-work'', ''mining''). Typically, this is to require the block hash to be less than a certain value. A limited block rate allows nodes to achieve eventual consensus, and a hard problem prevents attackers from ''overtaking'' the creation of legitimate blocks with fraudulent one. Hence, a successful attack requires control of $> 50\%$ of the total hashing power of all nodes.

The probability for a transaction in the main branch to become invalid due to a chain reorganization decreases with each block that is mined on top of it, but in principle it is always possible for a transaction to become invalid. In addition to this lack of finality of consensus, this approach induces significant latency as applications must wait not only for one block but many to be created. Applications that use the blockchain infrastructure must actively monitor the status of all transactions of interest, must react to chain reorganizations, and communicate this information to the user.

\subsection{BFT-Based Consensus and State Machine Replication}
\label{sec:bft}

In response to the drawbacks of the proof-of-work consensus provably correct ordering algorithms based on distributed systems research have seen a resurgence in interest. Most of the ongoing research can be traced back to a practical method for achieving byzantine fault tolerance (PBFT) \cite{DBLP:journals/tocs/CastroL02} where tolerating up to $f$ faulty nodes requires $3f+1$ total nodes. PBFT achieves consensus on the order of requests using a set of fully-connected ordering nodes. 

\paragraph*{Protocol}

Every ordering consensus is established by a specific set of nodes (''view''), with a leader or primary node. A client sends a request to all nodes. The leader proposes a sequence number for the request and broadcasts this in a \emph{pre-prepare} message. Upon receipt of a \emph{pre-prepare} message, a node broadcasts a corresponding \emph{prepare} message if it has itself received the request, and has not already received another pre-prepare message for the same sequence number. This indicates the node is prepared to accept the proposed sequence number. Nodes wait to receive $2f$ matching \emph{prepare} messages, indicating that $2f+1$ nodes are prepared to accept the proposed sequence number for the request. When a node has received $2f$ identical \emph{prepare} messages, it broadcasts a \emph{commit} message to all nodes. Nodes then waits to receive $2f$ identical \emph{commit} messages, indicating that $2f+1$ nodes have accepted the proposed sequence number for the request. Upon committing, the node executes the request and sends a \emph{reply} message to the client. The client waits for $2f+1$ identical replies, which indicates that a consensus has been reached on the sequence number of the request. 

The leader is not a fixed, central, or privileged node and is changed by consensus when nodes detect an unresponsive or malfunctioning leader. Leader change uses a three-stage protocol similar to the normal ordering protocol. 

Consensus request sequencing is closely related to state machine replication (SMR): Every node maintains a state that can be changed by client requests. When every node begins with the same state and executes requests in the same order, the state machine is replicated. 

\paragraph*{Implementation}

BFT-SMART \cite{DBLP:conf/dsn/BessaniSA14} is a software library built around a BFT protocol and adds dynamic view reconfiguration (nodes can join and leave views) and state exchange. BFT-SMART provides a simple programming interface. The client-side interface allows submission of requests. Applications implement a server-side interface, encapsulating the state machine, that receives ordered requests in consensus sequence from the library for execution. Requests are simple byte arrays and opaque to the library, the client- and server-side applications must serialize and deserialize these in a meaningful way. View reconfigurations (adding or removing a node, or changing the level of fault tolerance) are special types of requests but are treated as any other request for ordering and consensus purposes. For state exchange, the server-side application implements methods to fetch and set state snapshots, also serialized as byte arrays. When a node joins a view, it is sent the latest checkpointed state using collaborative state transfer, and requests after the checkpoint are then replayed, allowing the server state to catch up to the consensus state.

BFT-SMART has been proven to be correct and live, i.e. it will provide the same sequence of operations to all nodes and will not deadlock \cite{DBLP:conf/dsn/BessaniSA14}. In terms of throughput, a system with four nodes ($f=1$) has been shown to support more than 15,000 requests (1kB size) per second with latencies around 10 milliseconds on a local network. The performance decreases linearly as fault tolerance (and hence the number of nodes) increases: A system with 10 nodes ($f=3$) has been shown to support more than 10,000 requests per second \cite{DBLP:conf/dsn/BessaniSA14}.

\paragraph*{Summary}

BFT-based ordering avoids the latency, lack of finality and computational demands of proof-of-work consensus. On the other hand, its three-stage protocol imposes significant communication overhead and requires fully-connected nodes. Fault tolerance in BFT increases linearly with the number of nodes, but performance decreases due to additional communication. \emph{The different strengths and weaknesses of the two consensus mechanisms suggest that BFT-based ordering is a good fit with small, permissioned blockchains as they are used in the inter-organizational collaborative workflow management context.} This is echoed by \cite{ViriyasitavatHoonsopon2018blockchain}, who recommend BFT-based consensus for workflow execution because ''it guarantees safety, liveness, and some degree of fault tolerance'' and proof-of-work is ''impractical since the confirmation settlement is too long and unreliable''.

\section{General Architecture}
\label{sec:architecture}

The main component of a workflow management system is the workflow engine, which interprets the workflow specification and enables work items for execution by external services \cite{hollingsworth1995workflow}. Prior work (Sec.~\ref{sec:related}) has deployed the workflow engine on the blockchain itself, by compiling BPMN workflow specifications to smart contracts or by implementing a BPMN intepreter as a smart contract. In this paper, we treat the distributed blockchain as an infrastructure layer for existing off-chain workflow engines. We use the blockchain only for storing and sharing the state of work and achieving consensus on that state. To our knowledge, there has been no such implementation using BFT-based or any other ordering mechanism. 

Ordering service, block service, and the workflow engine are the three main components in our system architecture. In contrast to proof-of-work based blockchains, our architecture requires no mining service, no transaction service to manage pending transactions, and no virtual machine to execute or validate smart contract operations.

\paragraph*{Ordering Service} The ordering service in our prototype uses the BFT-SMART library \cite{DBLP:conf/dsn/BessaniSA14}. It consists of a client and a server. The ordering service client receives requests from clients and submits them to the ordering layer. Once ordered, the ordering layer submits the requests in consensus sequence to the ordering service server. The ordering layer maintains a fully connected network between all ordering nodes. Messages on this network are encrypted and signed using pre-distributed public/private keys.

\paragraph*{Block Service} The block service stores the blockchain, may exchange blocks with other nodes, and verifies the integrity of the blockchain. The block service uses a peer-to-peer network for block exchange with new and recovering nodes. This network is distinct from the ordering layer network and is not fully connected, but is encrypted and authenticated using the same public/private keys. Verification of the blockchain proceeds backwards from the block with the latest hash and any missing blocks are requested from other peers and verified prior to adding them to the local blockchain.

\paragraph*{Workflow Engine} The workflow engine maintains information about work items, workflow instances (cases), and workflow specifications. Together, this information forms the ''state of work'' or ''workflow state''. We call any operation that changes the workflow state a ''workflow operation''. The workflow engine interacts with services that provide resource management and worklists for user tasks, and with external services for service tasks.

The ordering service uses the term ''request'' to denote the objects it is ordering, the block services uses the term ''transaction'' to denote the objects it stores in blocks, and the workflow engine uses the term ''workflow operation'' to denote the objects that change the workflow state. In our architecture, these terms denote the same object: A workflow operation is ordered as a request, stored as a transaction on the blockchain, and executed by the workflow engine. We define workflow operations using lifecycle models for workflow specifications at runtime, workflow cases, and work items. Any transition in such a lifecycle model, such as the creation of a new specification or the completion of a work item, is a workflow operation. The XES standard \cite{acampora2017ieee} defines a work item lifecycle, as does the YAWL system \cite{ter2009modern}.

In principle, a system architecture can encompass different numbers of ordering services, block services, and workflow engines, distributed in different combinations on different network nodes. However, as the absence of trust among participating actors is a key motivation for the use of blockchains, we assume that every process participant requires and provides its own workflow engine, block service, and ordering service. We call this combination of workflow engine, block service, and ordering service a ''node'' in our architecture. 

This assumption significantly simplifies the architecture and implementation. Most importantly, new blocks can be created and stored locally on each node from ordered requests. In proof-of-work blockchains, new blocks are created by a single node, the successful mining node, and then distributed to other nodes. A more efficient alternative that is possible in our architecture is for every ordering service server to create new blocks from ordered requests and pass the new blocks directly to the local block service for inclusion in the blockchain. As the order of requests is identical for all nodes, the created blocks will be identical. This removes the need for block distribution, avoiding latencies and differences in block order.

In proof-of-work blockchains, blocks contain multiple transactions and mining nodes maintain a pool of pending transaction. The number of transactions in a block is a trade-off between desired transaction throughput, available hashing power, desired block creation rate, available network bandwidth, and tolerance for latency. In contrast, in BFT-based systems, there is no expensive mining. Hence, there is no reason to delay block creation and for blocks to contain multiple transactions: The blockchain becomes a chain of transactions.

\section{Adapting YAWL}
\label{sec:prototype}

The YAWL (''Yet Another Workflow Language'') workflow management system \cite{ter2009modern} is an open source workflow system for the YAWL language. The YAWL language is based on workflow nets, with significant extensions, and was designed to allow specification of workflow patterns \cite{van2003workflow}. The YAWL system consists of a workflow engine, a resource service, a workflow specification editor (modeling component), and a set of other services. The YAWL engine maintains workflow specifications for run-time, workflow case information and work item information. The resource service maintains the organizational model, work item allocations and provides worklist management. The resource service also provides the graphical user interfaces for user tasks. The specification editor is a Java application. The engine and other services are implemented as a set of web applications for a Java application server (typically Apache Tomcat). These provide a number of Java servlet APIs for communication between the engine and different services. Communication is done with XML documents; persistence is managed through a Hibernate layer in a relational database (typically PostgreSQL). 

YAWL separates resource management of work items from the work item lifecycle that is relevant to the case progress. For example, after the workflow engine creates (enables) a work item, the resource service manages the work item in the worklists of the resources it has been offered or allocated to. Offering, allocation, de-allocation, re-offering, etc. is not relevant to the workflow engine and consequently the engine is not involved or notified. Hence, we do not consider these operaitons as workflow operations. When a resource starts work on a work item, the resource service notifies the engine of this operation (''check-out''), as this operation is relevant to the case progression, e.g. for a deferred choice pattern, because it fires the associated workflow net transition. Hence, we consider this a workflow operation. Any subsequent resource changes such as delegation or re-assignment are managed internally by the resource service without involvement of the engine. These operations are not workflow operations. Only when the work item is completed, is the engine notified (''check-in''). This is another workflow operation.

Our adaptation of YAWL is guided by these principles:
\begin{enumerate}
\item Every organization collaborating in the inter-organizational workflow provides its own block service, ordering service, YAWL workflow engine and associated YAWL services. Together, these form a ''node'' (cf. Sec~\ref{sec:architecture}).
\item Each task in a workflow specification is assigned to a single node.
\item Organizational resources and their unique identifiers are local to each node.
\item Resource management of work items is performed locally for each organization/node and is only locally relevant. This reflects the strict separation between workflow-relevant work item changes and the resource-relevant work item changes in YAWL.
\end{enumerate}

Because the resource management is local to each node, the YAWL resource service requires no adaptations. The following subsections desribe primarily the adaptations to YAWL workflow engine.

\subsection{Workflow Specifications}

The only adaptation required for workflow specifications is the addition of a node identifier for each task. We use the BFT-SMART node identifier for this purpose, which is a simple integer value that is mapped to IP addresses through the BFT-SMART configuration. We updated the XML schema and the schema processors, and extended the YAWL editor to allow designers to specify this property for each atomic task.

\subsection{Engine Adaptations}

\begin{figure}
\centering
\includegraphics[width=0.5\columnwidth]{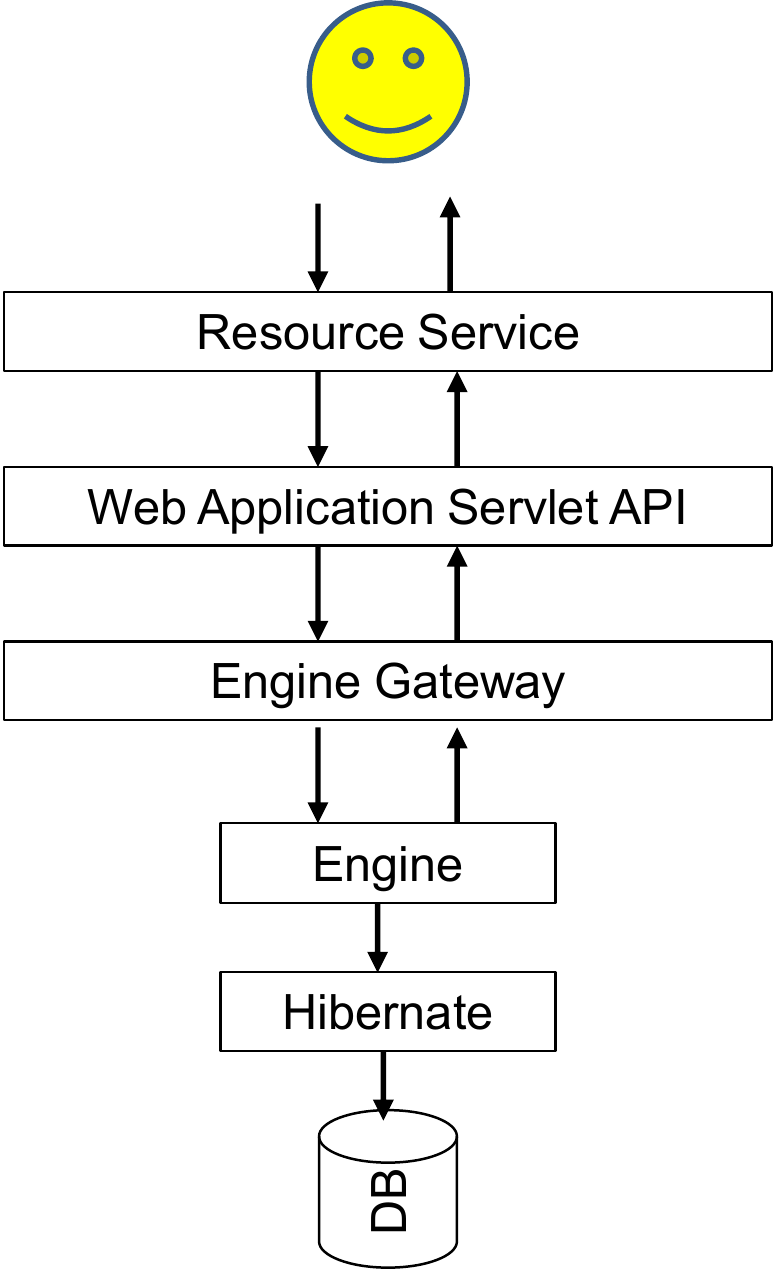}
\caption{YAWL architecture (as-is)}
\label{fig:yawl_architecture}
\end{figure}

The YAWL workflow engine is a singleton Java class (YEngine). Outside services do not interact with the engine directly but through an encapsulating singleton class called the engine gateway (YEngineGateway). This in turn is accessed from a number of HTTP servlets that expose various interfaces of the engine. Interface A is an administrative interface to manage external services and their authentication mechanisms. Interface B is the main interface for workflow aspects, such as launching and cancelling cases, starting and completing work items, etc. Interface E provides information for logging purposes. Interface X allows exception handlers to interact with the engine. Figure~\ref{fig:yawl_architecture} shows this architecture where all four interfaces are jointly represented by the ''Web Application Servlet API''. The workflow engine announces changes to the workflow state, such as new workitems or workitems skipped or cancelled due to expired timers, to registered external services through outgoing HTTP requests.

To adapt the YAWL engine to a blockchain infrastructure, we intercept inbound calls to the engine gateway to order them, distribute them to all nodes, and include them in the blockchain. They are then passed back to the engine gateway and the workflow engine for execution. We use the BFT-SMART library for ordering (cf. Sec.~\ref{sec:architecture}). We essentially split the engine gateway into a client-side and a server-side part and sandwich the ordering and block services in between. Our adapted architecture is shown in Fig.~\ref{fig:yawl_blockchain_architecture}. Also shown in the figure are the steps to process a workflow request, explained in detail below.

Calls to the engine gateway can be categorized on two dimensions. First, some calls are read-only, while others change the state of the engine. Second, some calls are local to each engine while others are globally relevant. Table~\ref{tab:calltypes} shows these categories and provides examples. Of all engine gateway calls, only the globally relevant ones need to be intercepted; the local ones are passed on directly to the local engine. 

\begin{table}
\centering
\begin{tabular}{|l|c|c|} \hline
& {\bf Read} & {\bf Write} \\ \hline
{\bf Local} & Retrieve services & Register new service \\ \hline
{\bf Global} & Retrieve case data for work item & Launch case \\ \hline
\end{tabular}
\caption{Workflow engine access types and examples}
\label{tab:calltypes}
\end{table}

While the majority of workflow operations that need ordering and must be captured on the blockchains are concerned with the work item lifecycle, some apply to workflow specifications and cases. Table~\ref{tab:interceptedwritecalls} provides a list of the intercepted write/update workflow engine calls. Table~\ref{tab:interceptedreadcalls} shows a list of the intercepted read workflow engine calls.

\begin{table}
\centering
\begin{tabular}{|l|} \hline
{\bf Workflow Specifications} \\ \hline
loadSpecification \\ \hline
unloadSpecification \\ \hline
{\bf Workflow Cases} \\ \hline
launchCase \\ \hline
cancelCase \\ \hline
{\bf Work Items} \\ \hline
suspendWorkItem \\ \hline
unsuspendWorkItem \\ \hline
rollbackWorkItem \\ \hline
completeWorkItem \\ \hline
startWorkItem \\ \hline
skipWorkItem \\ \hline
createNewInstance \\ \hline
restartWorkItem \\ \hline
cancelWorkItem \\ \hline
rejectAnnouncedEnabledTask \\ \hline
\end{tabular}
\caption{Intercepted and ordered write/update workflow operations}
\label{tab:interceptedwritecalls}
\end{table}

\begin{table}
\centering
\begin{tabular}{|l|} \hline
{\bf Workflow Specifications} \\ \hline
getProcessDefinition \\ \hline
getSpecificationDataSchema \\ \hline
getStartingDataSnapshot \\ \hline
getSpecificationList \\ \hline
getSpecificationData \\ \hline
getLatestSpecVersion \\ \hline
{\bf Workflow Cases} \\ \hline
getCasesForSpecification \\ \hline
getSpecificationIDForCase \\ \hline
getSpecificationForCase \\ \hline
getAllRunningCases \\ \hline
getCaseState \\ \hline
getCaseData \\ \hline
getCaseInstanceSummary \\ \hline
exportCaseState \\ \hline
exportAllCaseStates \\ \hline
{\bf Work Items} \\ \hline
getAvailableWorkItemIDs \\ \hline
getWorkItem \\ \hline
describeAllWorkItems \\ \hline
getWorkItemsWithIdentifier \\ \hline
getWorkItemsForService \\ \hline
getTaskInformation \\ \hline
checkEligibilityToAddInstance \\ \hline
getChildrenOfWorkItem \\ \hline
getWorkItemOptions \\ \hline
getMITaskAttributes \\ \hline
getResourcingSpecs \\ \hline
getWorkItemInstanceSummary \\ \hline
getParameterInstanceSummary \\ \hline
\end{tabular}
\caption{Intercepted and ordered write/update workflow operations}
\label{tab:interceptedreadcalls}
\end{table}

\paragraph*{Write Requests}

We intercept the globally relevant write or update calls for ordering, distribution, and inclusion in the blockchain. They are submitted as requests to the ordering service client, ordered, and handled by each ordering service server. The ordering service server creates new blocks for the block services, which in turn provides transactions in new blocks to the engine gateway for execution by the workflow engine. Any result from the workflow engine is returned by each ordering service server, in addition to the latest block hash, to the ordering layer. The ordering layer returns the consensus result to the ordering service client or signals request failure when no consensus is achieved.The path of write/update workflow engine calls that require ordering through the various components is as follows:

\begin{figure}
\centering
\includegraphics[width=\columnwidth]{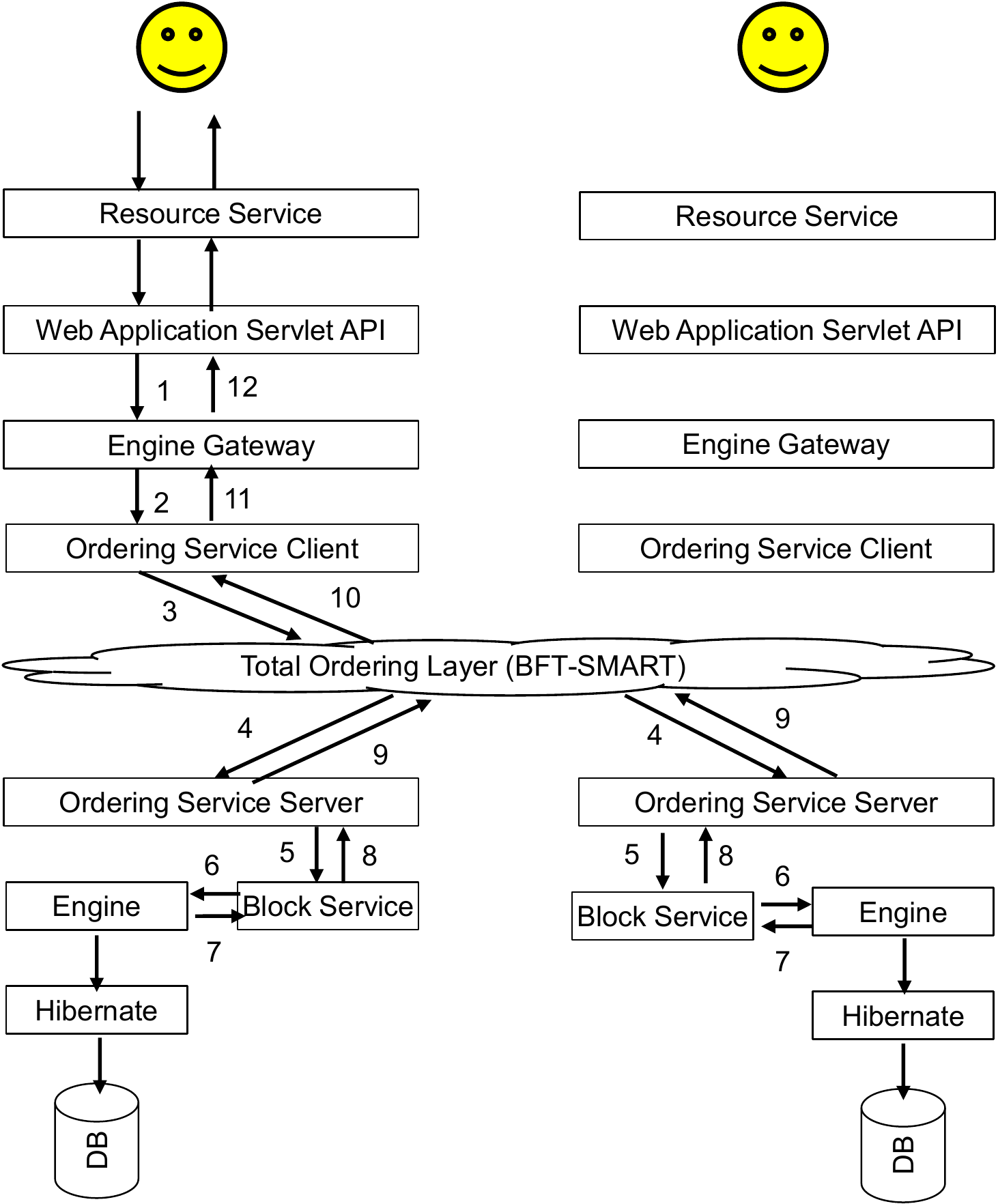}
\caption{YAWL on blockchain architecture and workflow event processing}
\label{fig:yawl_blockchain_architecture}
\end{figure}

\begin{enumerate}
\item Call received as HTTP POST method by servlet, submitted to engine gateway
\item Engine gateway identifies call that requires ordering, submits call as request to ordering service client
\item Ordering service client submits request to ordering layer
\item Ordering layer passes ordering requests to \emph{all} ordering service servers
\item Every ordering service server passes request to local block service for inclusion as a transaction in blockchain
\item Every block service passes the new transaction to local engine for processing, using original engine API
\item Engine returns result to blockservice
\item Block service returns engine result with latest block hash to ordering service server
\item Each ordering service server returns local result to ordering layer
\item Ordering layer returns consensus result to ordering service client or indicates lack of consensus.
\item Ordering service client returns result to engine gateway
\item Engine gateway returns result to HTTP servlet to be returned as HTTP response
\end{enumerate}

\paragraph*{Read Requests}

There are three options for handling inbound read requests for globally relevant data. Read requests do not need to be recorded on the blockchain. The first option is to submit a read request straight to the local workflow engine, bypassing the ordering layer (arrow '2a' in Fig.~\ref{fig:yawl_blockchain_architecture}). This is the fastest option but assumes a non-faulty local node. The second option is to submit a read request to the ordering layer as an unordered request, but for which a consensus result is provided from all ordering service servers. The third option is to submit a read request for ordering through the ordering service and obtaining the consensus result. The third option has the additional benefit of preventing inconsistent reads. Consider a situation where a read request is submitted to the ordering layer after a write request, but executed prior to the write request. From the perspective of the requestor, the result would be inconsistent. This inconsistency is avoided in the third option. However, this option is also the slowest. The second and third options provide the consensus workflow state to the caller even if the local node is faulty. They are depicted by arrows '5a' and '8a' in Fig.~\ref{fig:yawl_blockchain_architecture}. For these options, the path of read requests is the same as for write requests, except that they bypass the block service as no new block is created. As each option has advantages and disadvantages our system can be configured to provide any of them.

\subsection{Fault Detection and Recovery}
\label{sec:fault}

When both write requests and read requests are intercepted and ordered for consensus, the external services that access a node receive consensus information even if the local block service or workflow engine are faulty or unresponsive. This is, after all, the goal of byzantine fault tolerance. In effect, these services access a redundant, distributed, fualt-tolerant workflow management system through the ordering service client.

As shown in Fig.~\ref{fig:yawl_blockchain_architecture}, the client-side engine gateway has direct access to the local workflow server (arrow '2a'). Hence, it is possible for a node to identify when its local result for a read request differs from the consensus result. For this, our system can submit the same request along arrow '2a' as well as '2' and compare the results. This cannot be done for write requests so as not to doubly execute them. However, many write operations in YAWL return the description of the changed entity, such as the work item or the case identifier. Our system can retrieve the local copy of the entity via arrow '2a' and compare that against the result of the write operation issued along arrow '2'.

Differences between local and consensus state may indicate a faulty local block service or workflow engine. However, in fast-moving processes, differences might also arise due to timing. Because the consensus mechanism introduces a delay (even though it is very brief, on the order of milliseconds), the local state is read at a different time than the state for the consensus finding (arrows '2a' and '5a' in Fig.~\ref{fig:yawl_blockchain_architecture}). In that case, differences do not indicate a local fault.

While external services accessing a node can continue to function because they receive consensus information, differences between local and consensus information can indicate when to reset and recover a node. Whether and when nodes should be recovered depends the desired fault tolerance, to what extent the local node owner is willing to accept faulty information in the local blockchain or workflow engine, and on whether the differences are due to real workflow evoluation in fast-moving processes. Our system can be configured to fail early. When differences between local and consensus information are detected, the node is reset and recovered. Our system can also be configured to keep operating; in this case there is no comparison between local and consensus state but the node operator can manually issue a command to reset and recover. Node recovery deletes the local blockchain replica, the local ordering service server state, all information in the local workflow engine database, and then re-initializing the node as if it was a new node entering the system (Sec.~\ref{sec:startup}). 

\subsection{Time}

Because nodes do not generally have synchronized clocks, time-dependent requests may be problematic. Even synchronized clocks do not guarantee that requests are executed by the workflow engines on different nodes at precisely the same time. Due to network latencies, computational load and other factors, requests reach the different nodes' workflow engines at slightly different times and are executed at different times.

In YAWL, time-dependency arises in three contexts. First, work items contain enablement, firing , and start times set by the local server on enablement, firing and start. Second, YAWL provides timer tasks. Third, YAWL provides the option of delayed (scheduled) case launches, where users specify a point in time for the case launch or a duration for the delay.

As noted above, many workflow operations, such as starting, skipping, cancelling or describing (retrieving) a work item, return the work item serialized as XML. The timing information in this serialization is necessarily local to each workflow engine. Hence, when such requests are run through the ordering service, no consensus can be established. Our system addresses this issue by omitting time information from the XML serialization for ordering consensus (arrows '3' to '10' in Fig.~\ref{fig:yawl_blockchain_architecture}). After the ordering service returns a consensus result, the local node compare its local work item information with the consensus result (arrow '2a' in Fig.~\ref{fig:yawl_blockchain_architecture}), and, if they match, substitutates the complete information from the local engine. This prevents loss of any information that may be required by external services, such as the resource service. If the local and consensus information do not match, depending on system configuration, an error is signalled to trigger node recovery or the consensus information is returned without the timing information (cf. Sec.~\ref{sec:fault}). 

Timer tasks in YAWL are managed through timers in the workflow engine. When a work item timer expires, the controlling YWorkItemTimer object skips the work item if it has been started but not yet completed (timer begins at work item start), or cancels the work item, if it has not been started yet (timer begins at work item enablement). When a work item is completed normally prior to timer expiry, the timer is simply cancelled by the workflow engine. To allow this mechanism to function, and the skipping or cancellation be captured by the ordering service and blockchain, we changed the YWorkItemTimer object to call the engine gateway, instead of the engine, even though it is controlled by the engine.

YAWL manages delayed (scheduled) case launches similarly to timer tasks. A YLaunchDelayer object is created that calls the engine to launch the new case upon expiry. Similar to our change to the work item timer, we adapted the YLaunchDelayer to call the engine gateway, instead of the engine.

\subsection{Work Item Visibility}

Enabled work items are announced to external services by the engine through an announcer object, to which external services register themselves as an observer. To limit work item visibility to the node they are associated with, we modified the announcer to check each announcement whether it is relevant to the local node: Announcements referring to cases are relevant to all nodes and are always announced; announcements referring to work items are relevant to specific nodes only and are announced only if they refer to the local node. Table~\ref{tab:announcements} shows a list of announcements, categorized by whether they are locally or globally relevant.

\begin{table}
\centering

\begin{tabular}{|l|} \hline
{\bf Global} \\ \hline 
Case start \\ \hline
Case completion \\ \hline
Case cancellation \\ \hline
Case deadlock \\ \hline
Case suspension \\ \hline
Case resumption \\ \hline
{\bf Local} \\ \hline
Work item firing \\ \hline
Work item status change \\ \hline
Work item cancellation \\ \hline
Timer expiry \\ \hline
\end{tabular}
\caption{Announced events, local events are announced only if they concern the local node}
\label{tab:announcements}
\end{table}

Additionally, write requests relevant to work items (Table~\ref{tab:interceptedwritecalls}) are only accepted if they refer to a work item that is assigned to the local node. This ensures only the local services can act on local work items and is necessary as a node's external service may have information also about non-local work items through the read requests in Table~\ref{tab:interceptedreadcalls}. This is checked by the engine gateway before requests are submitted to the ordering service (arrow '2' in Fig.~\ref{fig:yawl_blockchain_architecture}). This is also checked before workflow operations received from the block service are passed on to the engine for execution (arrow '6' in Fig.~\ref{fig:yawl_blockchain_architecture}). The latter check uses the originating node identifier in every transaction and is necessary because faulty or malicious nodes may submit a workflow operation that is not assigned to their node. Correctly operating nodes must identify and reject such an operation.

\subsection{Blockchain Monitor Service}

To allow for monitoring the ordering and block services, we implemented an additional web service. Through this service, it is possible to examine the blockchain with its blocks and transactions, to see the current state of the ordering service, i.e. the level of fault tolerance and the set of nodes that participate in the current ordering view, and it is possible to manually trigger reset and recovery of the node.  

\subsection{Node Startup}
\label{sec:startup}

The YAWL workflow engine database has been extended to store the hash of the last block that has been passed to it from the block service after executing each workflow operation. The last block hash is required during node startup, which proceeds as follows. The servlet container starts the interface B servlet, which in turn creates the singleton instance of the engine gateway. The engine gateway creates the singleton instance of the engine. During initialization, the engine reads the latest block hash from its database. Next, the ordering service server is started. The peer-to-peer network for block exchange is then started and contacts other peers. The first contacted peer initiates an ordering servie view change to join the new node into the BFT-SMART ordering view. As part of the BFT-SMART state exchange protocol, the new node receives the consensus last block hash. Then, the local ordering service client is started. The block service starts and, through the ordering service client, requests the latest block hash from the ordering service server. Comparing the local blockchain, the last block hash of the engine and the consensus block hash, it identifies and requests any missing blocks. Once all requested blocks are received, the block service validates the blockchain. The engine gateway then requests blocks from the engine's last block hash from the block service and replays these on the engine until the engine is caught up with the consensus blockchain. At this point the node is fully caught up with the consensus blockchain and has completed its initialization.

\section{Discussion}
\label{sec:discussion}

From the user's perspective, the set of nodes that form the distributed, fault-tolerant workflow system, look little different from a singluar YAWL system. In proof-of-work systems, users and workflow engines must be aware of and react to possible transaction invalidation, blockchain reorganization, eventual/delayed consensus and transactions pending their required ''assumed safe'' mining depth \cite{falazi2019modeling,falazi2019process}. In contrast, because of immediate and final consensus in the BFT protocols, our system behaves similar to non-blockchain systems, with no pending transactions or latency for block mining. The status of workflow operations cannot change and need not be monitored or reported to the user. Response times to the user are not noticeably longer than for a traditional YAWL system. 

\subsection{Correctness}

Systems that deploy the workflow engine on the blockchain as a smart contract enforce workflow consensus for every node as the workflow state is represented by the smart contract state. Submitting an illegal workflow operation, e.g. starting an already checked-out work item, by a faulty or malicious workflow engine will cause the smart contract to retain the legal state and dismiss the submitted workflow operation. If instead the faulty or malicious node attempts to submit a transaction with an invalid smart contract state, the miners will independently detect this during validation and ignore the transaction. Invalid transactions will also be detected by each node when blocks are received and validated.

In contrast, our approach guarantees that the majority of nodes will arrive at a consensus about the current workflow state (BFT approaches can tolerate up to $1/3$ malicious nodes). When a workflow engine submits an illegal workflow operation, e.g. to start an already checked out work item, the workflow engines of the non-faulty majority nodes will each, individually and separately, reject this operation and return a suitable error response. Hence, the consensus response signals that the work item state has not changed. This indicates to the requester that it is faulty. The requester can then be reset and recovered. Note that illegal requests are still ordered and are also stored in the blockchain. A limitation of our approach is that faulty nodes can only detect their own failure once they submit a transaction or request workflow information and the consensus result differs from their local result. They cannot detect their own faults while only receiving transactions in new blocks because the ordering service servers do not receive consensus results. 

When a node needs to catch up with the blockchain, the BFT-SMART state replication ensures that it receives the consensus last hash as new state from the set of running ordering nodes. With this, it is able to detect and reject incoming bad blocks as they are transferred. In general, assuming that there is a valid consensus on the ordering state (i.e. the last block hash), a node can always verify its blockchain and, if required, rebuild it by requesting blocks from other nodes. 

They key difference between proof-of-work approaches and ours is that blocks are not created on a single node, but on every node separately and concurrently. That is, the challenge is not to identify and reject bad (malicious) blocks as they are transferred, but only to ensure consensus ordering. The assumption of a majority of non-faulty nodes then ensures a majority of nodes with the correct workflow state.

\subsection{Limitations}
\label{sec:limitations}

\paragraph*{Workflow Engine Recovery}

Because the YAWL workflow engine was not designed for use on blockchain infrastructure, or any other kind of distributed infrastructure, it lacks a number of useful features. It does not have a checkpointing feature that would allow the engine to revert to a particular state, e.g. states designated by particular block hashes. It also lacks a rollback feature for workflow operations. Together this means that, when a node encounters an illegal, or non-concensus, state after submitting a workflow transaction, it can neither undo a number of transactions nor can it revert to a named checkpoint state. Instead, it must rebuild the worklow engine state by means of re-playing all transactions in the blockchain, which can be a very expensive operation.

\paragraph*{Resource Assignments}

There may be situations where it is desirable to specify that a work item may be performed by some resource role irrespective of the participating organization. However, as work items are assigned to a single node in our system, it is not possible to offer work items for execution to a set of resources across a set of collaborating organizations. This limits the extent of collaboration in the workflow. Similarly, delegetion or re-assignment of work items are also not possible across organizational boundaries. At the organizational level, this limitation can only be overcome by a global organizational model, which requires participating organizations to agree on role or skill definitions that are relevant to resource management. This is a non-trivial challenge especially in the absence of trust. At the technical level, this requires moving not only the workflow engine but also the resource service to a distributed model using the blockchain. This too is a non-trivial challenge.

\paragraph*{Resource Specifications}

A YAWL workflow specification contains resource identifiers for assigning each task's resourcing. One of the limitations of the YAWL specification editor is that it connects to a single YAWL resource service to access organizational data for use in workflow specifications. Because nodes maintain their own resource information, constructing a global workflow specification must be done by manually adding the necessary local resource identifiers to the workflow specification. 

\paragraph*{Throughput and Scalability}

While our approach has lower latency and higher throughput than proof-of-work blockchains, it does not scale to a very large number of nodes. Given these characteristics, our architecture is suitable for permissioned blockchain applications using a small group of participants (on the order of a few tens). The low latency and high throughput also make them suitable for fast-moving processes, where activities are of short duration or must follow each other quickly. For example, our transaction throughput time is well below one second, whereas many proof-of-work blockchains operate at latencies on the order of minutes.

\paragraph*{Resilience}

An often discussed attack on proof-of-work based blockchains requires a malicious actor to control the majority ($> 50\%$) of the total hashing power of all nodes. In contrast, attacking a BFT-based system requires control of more than $1/3$ of all nodes. Assuming equal hashing power for all nodes, the proof-of-work based blockchain appears more resilient to attacks. However, in many use cases, this assumption is unlikely to hold. Small networks and networks where a few actors control significant resources are particularly prone to an imbalance in hashing power. In contrast, attacking a BFT-based system cannot be done by concentrating computational power but requires control of more than $1/3$ of all nodes, which is difficult to achieve in the absence of trust among actors. As a result, resilience to attacks and faults cannot be easily compared between proof-of-work and BFT-based blockchains; it is context and application dependent. 

\paragraph*{Workflow, Trust, and Fault Tolerance Requirements}

In our approach, the number of nodes must strike a balance between the requirements of the workflow, the level of fault tolerance, and the performance of the system. The number of ordering nodes is determined by the desired level of fault tolerance, whereas the number of workflow nodes is determined based on the use case and the number of participating actors. A use case requiring more ordering than workflow nodes (e.g. because some actors share a workflow engine but do not wish to relinquish control over the blockchain infrastructure) can be accommodated by nodes that are not assigned any workflow activities. On the other hand, when a use case requires more workflow nodes than ordering nodes (e.g. because groups of actors trust each other), the excess ordering nodes decrease performance due to the BFT protocol communication overhead. This drawback can only be addressed by relaxing the trust requirements, i.e. groups of actors must partially trust each other, so that the $1:1$ correspondence between ordering service, block service, and workflow engine can be relaxed. 

\subsection{Future Extensions}

As our system works at the level of work items and their lifecycle transitions, rather than the semantics of a workflow specification language such as BPMN, an extension to heterogeneous workflow engines is readily possible. This requires a mapping of work item lifecycles of the different workflow engines, possibly with a canonical intermediate lifecycle model to be used on the ordering service layers. When workflow requests are intercepted, they are translated to the canonical lifecycle interchange model, ordered, and back-translated to particular engine lifecycle models and workflow operations by each ordering service server prior to execution by the local workflow engine. For our own future work, we are particularly interested in other open-source workflow systems, such as the Bonita system, due to their easy adaptability.

Previous work on blockchain-based WfMS has focused on smart contracts and proof-of-work based blockchains. However, proof-of-work-based systems have significant drawbacks in terms of processing power requirements, latency, and the lack of final consensus. In this work, we have shown that a BFT-derived ordering and consensus method is a suitable WfMS infrastructure. Even without the use of smart contracts, the use of a blockchain remains essential, as it provides independent validation of workflow actions, distribution, replication, and tamper-proofing to workflow management systems.

While there are limitations to the BFT-based approach (cf. Sec.~\ref{sec:limitations}), our approach also has significant advantages over proof-of-work based approaches:

\begin{itemize}
\item Our system is cheaper to operate than public proof-of-work blockchains that incentivize block mining through cryptocurrencies. While proof-of-work based blockchains may be deployed privately, they are then open to increased risk of attack (cf. Sec.~\ref{sec:limitations}).
\item Our system provides immediate and final consensus. This means that from both the workflow modeller's perspective and the user's perspective, the system looks and behaves like a traditional workflow engine. Neither the workflow designer nor the user need to deal with issues of transaction status or eventual transaction invalidation.
\item Our system provides a greater throughput than proof-of-work based approaches. 
\item Not relying on smart contracts enables porting of existing feature-complete workflow engines to blockchain infrastructure. This allows rich workflow languages and leverages existing implementations. \end{itemize}

To conclude, this paper has presented a prototype implementation for an architecture that has not yet seen any attention in the blockchain-based workflow literature. We have implemented a BFT-based system as recommended by \cite{ViriyasitavatHoonsopon2018blockchain} and shown that this infrastructure is suitable as the infrastructure foundation for adapting existing workflow management systems to support inter-organizational workflows.

\bibliographystyle{IEEEtran}

\begin{thebibliography}{10}
\providecommand{\url}[1]{#1}
\csname url@samestyle\endcsname
\providecommand{\newblock}{\relax}
\providecommand{\bibinfo}[2]{#2}
\providecommand{\BIBentrySTDinterwordspacing}{\spaceskip=0pt\relax}
\providecommand{\BIBentryALTinterwordstretchfactor}{4}
\providecommand{\BIBentryALTinterwordspacing}{\spaceskip=\fontdimen2\font plus
\BIBentryALTinterwordstretchfactor\fontdimen3\font minus
  \fontdimen4\font\relax}
\providecommand{\BIBforeignlanguage}[2]{{%
\expandafter\ifx\csname l@#1\endcsname\relax
\typeout{** WARNING: IEEEtran.bst: No hyphenation pattern has been}%
\typeout{** loaded for the language `#1'. Using the pattern for}%
\typeout{** the default language instead.}%
\else
\language=\csname l@#1\endcsname
\fi
#2}}
\providecommand{\BIBdecl}{\relax}
\BIBdecl

\bibitem{ter2009modern}
\BIBentryALTinterwordspacing
A.~H.~M. ter Hofstede, W.~M.~P. van~der Aalst, M.~Adams, and N.~Russell, Eds.,
  \emph{Modern Business Process Automation - {YAWL} and its Support
  Environment}.\hskip 1em plus 0.5em minus 0.4em\relax Springer, 2010.
  [Online]. Available: \url{http://www.yawlbook.com/home/}
\BIBentrySTDinterwordspacing

\bibitem{fridgen2017implementation}
\BIBentryALTinterwordspacing
G.~Fridgen, B.~Sablowsky, and N.~Urbach, ``Implementation of a blockchain
  workflow management prototype,'' \emph{{ERCIM} News}, vol. 2017, no. 110,
  2017. [Online]. Available:
  \url{https://ercim-news.ercim.eu/en110/special/implementation-of-a-blockchain-workflow-management-prototype}
\BIBentrySTDinterwordspacing

\bibitem{fridgen2018cross}
\BIBentryALTinterwordspacing
G.~Fridgen, S.~Radszuwill, N.~Urbach, and L.~Utz, ``Cross-organizational
  workflow management using blockchain technology - towards applicability,
  auditability, and automation,'' in \emph{51st Hawaii International Conference
  on System Sciences {HICSS}}.\hskip 1em plus 0.5em minus 0.4em\relax {AIS}
  Electronic Library, 2018. [Online]. Available:
  \url{http://aisel.aisnet.org/hicss-51/in/blockchain/6}
\BIBentrySTDinterwordspacing

\bibitem{hukkinen2017distributed}
T.~Hukkinen, J.~Mattila, T.~Sepp{\"a}l{\"a} \emph{et~al.}, ``Distributed
  workflow management with smart contracts,'' The Research Institute of the
  Finnish Economy, Tech. Rep., 2017.

\bibitem{weber2016using}
I.~Weber, X.~Xu, R.~Riveret, G.~Governatori, A.~Ponomarev, and J.~Mendling,
  ``Using blockchain to enable untrusted business process monitoring and
  execution,'' Technical Report UNSW-CSE-TR-201609, University of New South
  Wales, Tech. Rep., 2016.

\bibitem{weber2016untrusted}
\BIBentryALTinterwordspacing
------, ``Untrusted business process monitoring and execution using
  blockchain,'' in \emph{Business Process Management - 14th International
  Conference, {BPM}, Proceedings}, ser. Lecture Notes in Computer Science,
  M.~L. Rosa, P.~Loos, and O.~Pastor, Eds., vol. 9850.\hskip 1em plus 0.5em
  minus 0.4em\relax Springer, 2016, pp. 329--347. [Online]. Available:
  \url{https://doi.org/10.1007/978-3-319-45348-4\_19}
\BIBentrySTDinterwordspacing

\bibitem{rimba2017comparing}
\BIBentryALTinterwordspacing
P.~Rimba, A.~B. Tran, I.~Weber, M.~Staples, A.~Ponomarev, and X.~Xu,
  ``Comparing blockchain and cloud services for business process execution,''
  in \emph{2017 {IEEE} International Conference on Software Architecture,
  {ICSA}}.\hskip 1em plus 0.5em minus 0.4em\relax {IEEE} Computer Society,
  2017, pp. 257--260. [Online]. Available:
  \url{https://doi.org/10.1109/ICSA.2017.44}
\BIBentrySTDinterwordspacing

\bibitem{rimba2018quantifying}
------, ``Quantifying the cost of distrust: Comparing blockchain and cloud
  services for business process execution,'' \emph{Information Systems
  Frontiers}, pp. 1--19, 2018.

\bibitem{garcia2017optimized}
\BIBentryALTinterwordspacing
L.~Garc{\'\i}a-Ba{\~n}uelos, A.~Ponomarev, M.~Dumas, and I.~Weber, ``Optimized
  execution of business processes on blockchain,'' in \emph{Business Process
  Management - 15th International Conference, {BPM}, Proceedings}, ser. Lecture
  Notes in Computer Science, J.~Carmona, G.~Engels, and A.~Kumar, Eds., vol.
  10445.\hskip 1em plus 0.5em minus 0.4em\relax Springer, 2017, pp. 130--146.
  [Online]. Available: \url{https://doi.org/10.1007/978-3-319-65000-5\_8}
\BIBentrySTDinterwordspacing

\bibitem{dijkman2008semantics}
\BIBentryALTinterwordspacing
R.~M. Dijkman, M.~Dumas, and C.~Ouyang, ``Semantics and analysis of business
  process models in {BPMN},'' \emph{Information {\&} Software Technology},
  vol.~50, no.~12, pp. 1281--1294, 2008. [Online]. Available:
  \url{https://doi.org/10.1016/j.infsof.2008.02.006}
\BIBentrySTDinterwordspacing

\bibitem{lopez2017caterpillar}
\BIBentryALTinterwordspacing
O.~L{\'{o}}pez{-}Pintado, L.~Garc{\'{\i}}a{-}Ba{\~{n}}uelos, M.~Dumas, and
  I.~Weber, ``Caterpillar: {A} blockchain-based business process management
  system,'' in \emph{Proceedings of the {BPM} Demo Track co-located with 15th
  International Conference on Business Process Modeling}, ser. {CEUR} Workshop
  Proceedings, R.~Claris{\'{o}}, H.~Leopold, J.~Mendling, W.~M.~P. van~der
  Aalst, A.~Kumar, B.~T. Pentland, and M.~Weske, Eds., vol. 1920, 2017.
  [Online]. Available:
  \url{http://ceur-ws.org/Vol-1920/BPM\_2017\_paper\_199.pdf}
\BIBentrySTDinterwordspacing

\bibitem{DBLP:journals/insk/CiccioCDGLLMPTW19}
\BIBentryALTinterwordspacing
C.~D. Ciccio, A.~Cecconi, M.~Dumas, L.~Garcia-Banuelos, O.~Lopez-Pintado,
  Q.~Lu, J.~Mendling, A.~Ponomarev, A.~B. Tran, and I.~Weber, ``Blockchain
  support for collaborative business processes,'' \emph{Informatik Spektrum},
  vol.~42, no.~3, pp. 182--190, 2019. [Online]. Available:
  \url{https://doi.org/10.1007/s00287-019-01178-x}
\BIBentrySTDinterwordspacing

\bibitem{STURM201919}
\BIBentryALTinterwordspacing
C.~Sturm, J.~Scalanczi, S.~Sch{\"{o}}nig, and S.~Jablonski, ``A
  blockchain-based and resource-aware process execution engine,'' \emph{Future
  Generation Computer Systems}, vol. 100, pp. 19 -- 34, 2019. [Online].
  Available:
  \url{http://www.sciencedirect.com/science/article/pii/S0167739X18327158}
\BIBentrySTDinterwordspacing

\bibitem{madsen2018collaboration}
M.~F. Madsen, M.~Gaub, T.~H{\o}gnason, M.~E. Kirkbro, T.~Slaats, and S.~Debois,
  ``Collaboration among adversaries: distributed workflow execution on a
  blockchain,'' in \emph{2018 Symposium on Foundations and Applications of
  Blockchain}, 2018.

\bibitem{DBLP:journals/tocs/CastroL02}
\BIBentryALTinterwordspacing
M.~Castro and B.~Liskov, ``Practical byzantine fault tolerance and proactive
  recovery,'' \emph{{ACM} Trans. Comput. Syst.}, vol.~20, no.~4, pp. 398--461,
  2002. [Online]. Available: \url{https://doi.org/10.1145/571637.571640}
\BIBentrySTDinterwordspacing

\bibitem{DBLP:conf/dsn/BessaniSA14}
\BIBentryALTinterwordspacing
A.~N. Bessani, J.~Sousa, and E.~A.~P. Alchieri, ``State machine replication for
  the masses with {BFT-SMART},'' in \emph{44th Annual {IEEE/IFIP} International
  Conference on Dependable Systems and Networks, {DSN} 2014, Atlanta, GA, USA,
  June 23-26, 2014}.\hskip 1em plus 0.5em minus 0.4em\relax {IEEE} Computer
  Society, 2014, pp. 355--362. [Online]. Available:
  \url{https://doi.org/10.1109/DSN.2014.43}
\BIBentrySTDinterwordspacing

\bibitem{ViriyasitavatHoonsopon2018blockchain}
W.~Viriyasitavat and D.~Hoonsopon, ``Blockchain characteristics and consensus
  in modern business processes,'' \emph{Journal of Industrial Information
  Integration}, 2018, preprint.

\bibitem{hollingsworth1995workflow}
D.~Hollingsworth, ``The workflow reference model,'' Workflow Management
  Coalition, Tech. Rep., 1995.

\bibitem{acampora2017ieee}
G.~Acampora, A.~Vitiello, B.~Di~Stefano, W.~Van Der~Aalst, C.~G{\"u}nther, and
  E.~Verbeek, ``Ieee 1849tm: The xes standard,'' \emph{IEEE Computational
  Intelligence Magazine}, 2017.

\bibitem{van2003workflow}
\BIBentryALTinterwordspacing
W.~M.~P. van~der Aalst, A.~H.~M. ter Hofstede, B.~Kiepuszewski, and A.~P.
  Barros, ``Workflow patterns,'' \emph{Distributed and Parallel Databases},
  vol.~14, no.~1, pp. 5--51, 2003. [Online]. Available:
  \url{https://doi.org/10.1023/A:1022883727209}
\BIBentrySTDinterwordspacing

\bibitem{falazi2019modeling}
G.~Falazi, M.~Hahn, U.~Breitenb{\"u}cher, and F.~Leymann, ``Modeling and
  execution of blockchain-aware business processes,'' \emph{SICS
  Software-Intensive Cyber-Physical Systems}, vol.~34, no. 2-3, pp. 105--116,
  2019.

\bibitem{falazi2019process}
G.~Falazi, M.~Hahn, U.~Breitenb{\"u}cher, F.~Leymann, and V.~Yussupov,
  ``Process-based composition of permissioned and permissionless blockchain
  smart contracts,'' in \emph{2019 IEEE 23rd International Enterprise
  Distributed Object Computing Conference (EDOC)}.\hskip 1em plus 0.5em minus
  0.4em\relax IEEE, 2019, pp. 77--87.

\end{thebibliography}


\end{document}